\documentclass[journal]{IEEEtran}
\usepackage{caption}
\usepackage{subcaption}
\usepackage{cite}
\usepackage{hhline}
\usepackage[cmex10]{amsmath}
\usepackage{graphicx}
    \graphicspath{{./figs/}}
    \DeclareGraphicsExtensions{.pdf}
\usepackage{balance}
\usepackage{algorithmic}
\usepackage{dsfont}
\usepackage{hhline}
\usepackage[usenames]{color}
\usepackage{multirow}
\usepackage{amssymb}
\usepackage{array}
\usepackage{url}
\usepackage{xspace}
\usepackage{mathtools}

\usepackage[table,xcdraw]{xcolor}
\usepackage{blindtext}
\usepackage{diagbox}
\usepackage{cleveref}
\usepackage{tablefootnote}

\begin{document}
%
\title{Digital Contact Tracing: Large-scale Geolocation Data as an Alternative to Bluetooth-based Apps' Failure}
%
%
%

\author{\IEEEauthorblockN{
José González-Cabañas\IEEEauthorrefmark{1},
Ángel Cuevas\IEEEauthorrefmark{1}\IEEEauthorrefmark{2},
Rubén Cuevas\IEEEauthorrefmark{1}\IEEEauthorrefmark{2},
Martin Maier\IEEEauthorrefmark{3},
}

\IEEEauthorblockA{\IEEEauthorrefmark{1} Universidad Carlos III de Madrid, Spain}\\
\IEEEauthorblockA{\IEEEauthorrefmark{2}UC3M-Santander Big Data Institute, Spain}\\

\IEEEauthorblockA{\IEEEauthorrefmark{3}Institut National de la Recherche Scientifique, Canada}\\


}
\maketitle

\begin{abstract}
The currently deployed contact-tracing mobile apps have failed as an efficient solution in the context of the COVID-19 pandemic. None of them has managed to attract the number of active users required to achieve an efficient operation.  This urges the research community to re-open the debate and explore new avenues that lead to efficient contact-tracing solutions. This paper contributes to this debate with an alternative contact-tracing solution that leverages already available geolocation information owned by BigTech companies with very large penetration rates in most countries adopting contact-tracing mobile apps.  Moreover, our solution provides sufficient privacy guarantees to protect the identity of infected users as well as precluding Health Authorities from obtaining the contact graph from individuals.
\end{abstract}


%
\IEEEpeerreviewmaketitle

\section{Introduction}


There is growing evidence that any strategy to effectively fight COVID-19 requires an efficient tracing of all contacts of infected individuals. Recent studies conclude that manual tracing is not fast enough and recommend the use of digital contact tracing systems able to use large-scale location information \cite{Science_COVID19_mobileapp}. A key element of the success of a digital contact-tracing system is its adoption. 

Singapore was one of the first countries implementing a digital contact-tracing system in early 2020. They opted to implement a mobile app that uses bluetooth (BT) technology to identify when two users have been in close proximity. If one of those users is tested positive in COVID-19, the other one is identified as a potential contagion. 20\% of the population in Singapore installed the mobile app. But this was not enough. Indeed, a responsible from the Ministry of Health of Singapore stated that they would need three quarters of the citizens installing the app to make the digital contact-tracing strategy successful \cite{Singapore_fail}. 

Although it is not clear what is the adoption rate from which a BT contact-tracing app becomes efficient in controlling a pandemic, some preliminary studies suggest that to mitigate the pandemic an adoption by 60\% of the population in a country would be required \cite{Science_COVID19_mobileapp}\cite{hinch2020effective}. Some simulations studies show that if the adoption is below 20\% the benefit of a BT contact-tracing app is very small, but we can observe a significant impact with 40+\%  adoption rate \cite{hinch2020effective}. Note that, by adoption rate we refer to the rate of people actively using the app, rather than the number of installations. 

BT-based contact-tracing apps have a major problem. They are newly released and thus they need to achieve the required high adoption rate in a short period of time from scratch.  To the best of our knowledge, neither researchers nor public or private institutions have proposed a convincing strategy to achieve the required adoption rate. For the time being, it seems that the success of any BT contact-tracing app depends solely on the self-responsibility of people, and it has not been enough.

Despite of the described problem and the reported failure of Singapore's app, most western countries  (especially in Europe) have also opted for mobile apps using BT technology as their contact-tracing systems. In particular, most of these countries opted for using the Decentralized Privacy-Preserving Proximity Tracing (DP-3T) protocol \cite{DP-3T}.  The main design goal of DP-3T is to provide full-privacy guarantees. In particular, it aims at guaranteeing that the contact-tracing applications using this protocol cannot be misused in the future for privacy-intrusive practices such as advertising, or even, massive surveillance. 
To support Health Authorities willing to deploy contact-tracing apps, Google and Apple integrated in their mobile operative systems (OS) a version of the DP-3T protocol. The OS just records user encounters using BT and offers this information to the mobile app, which implements the algorithm to identify risk contacts. In spite of this effort, to the best of our knowledge, none of the existing contact-tracing apps has significantly contributed to mitigate the virus transmission so far. For instance, early data from the Swiss Health Authority indicates that just 12\% of infected individuals report their positive through the app \cite{Swiss_Covid_Efficiency}. In Spain this number shrinks to 2\%.

In addition, as scientific evidences about the airbone transmission of COVID-19 become almost irrefutable \cite{airbone_transmision}, another important limitation of existing BT contact-tracing apps arises. They are designed to identify short-distance contact between two individuals, i.e., less than 2 meters apart. However, airbone transmission implies that contagion between two persons at longer distances is possible. Hence, existing BT contact-tracing apps may miss an important fraction of contacts that should be identified as risk contacts.

Finally, solutions like DP-3T, designed with the main goal of offering full-privacy, present further important shortcomes in the fight of a pandemic. Some of them are: 1) Even if the adoption rate were high, they require infected users voluntarily declare their positive condition through the app, leaving a very important task such as the control of a pandemic in the hands of individuals' decision. For instance, an early study in Switzerland shows that 1/3 of the users of the app tested positive did not use the app to report their case \cite{Swiss_Covid_Efficiency}; 2) The performance and efficiency of the contact-tracing app cannot be assessed. Not even how many infected users have been detected through the app, as recognized by authors of the DP-3T protocol \cite{Swiss_Covid_Efficiency}; 3) They are unable to provide even aggregate (and not privacy invasive) context information, which might be of great value to improve our knowledge on the transmission patterns used by COVID-19 (or other viruses). For instance, in this paper we consider the following one: revealing aggregate statistics of the type of locations (restaurants, sport facilities, public transportation, hospitals, etc) infected users visited while they were contagious may be useful to identify statistical biases on specific type of locations that may reveal hotspots for the virus transmission.  

Given the described context, the main goal of this paper is to urge the research community to expand the definition of digital contact-tracing systems having in mind the following key elements: 1) the experience gained so far invites to avoid solutions that require to achieve massive adoption from scratch; 2) contact-tracing solutions must be designed considering airbone transmission distances (few meters) as reference; 3) guide the design of the solutions setting the \emph{efficiency} in fighting the pandemic (i.e., saving lives and mitigating the impact on the economy) as the primary goal instead of \emph{privacy}. Of course, the proposed solution should be compliant with the existing data protection and privacy laws in the country where it is deployed. 

In this paper, we propose an alternative digital contact-tracing system based on the three previous key elements as fundamental design principles:

\begin{enumerate}

\item \textbf{High adoption rate:} We propose to use real-time location information from (literally) billions of people around the world that is already available in databases of large BigTech companies like Facebook (FB), Google, Apple, etc. We refer to these players as Location Providers (LPs) in this paper. Some of these LPs, mainly Google and Facebook, have a very large rate of active users, over 50\%, in many western countries. 

\item \textbf{Contact identification in airbone transmission range:} To geolocate users at both outdoor \cite{GPSAccuracy} and indoor locations \cite{WifiIndoorGoogle} with an accuracy of few meters, these BigTech firms use a combination of techniques that rely on multiple signals including: GPS location information, WiFi SSIDs signal's power, cellular network signals, etc. 

\item \textbf{Legal and Ethical Requirements:} We are interested in performing contact-tracing  just for individuals who has been tested positive of COVID-19. The identity of infected individuals is sensitive information handled by the Health Authority (HA) of each country, which is also responsible for running the contact-tracing strategy. Therefore, the HA has the identity of infected individuals while the LP  has the data to perform the contact-tracing for those individuals.  We propose a system that allows running contact-tracing using LPs data on those individuals tested positive as reported by HAs.
Even the most restrictive data protection laws, like the GDPR \cite{GDPR}, explicitly provision exceptions in which personal data can be used to monitor epidemics and their spread (see GDPR Article 6 Recital 46 \cite{GDPR}). Sustained on this legal basis an agreement to perform an exchange of data between LPs and HAs might be possible. However, to provide higher privacy guarantees, in this paper we propose a simple architecture and communication protocol that enable the exchange of information between a LP and a HA significantly limiting  the ability of (1) HAs to obtain the contact graph of an individual and (2) LPs to learn the identity of infected individuals. 

\end{enumerate}


We acknowledge that we have no evidence of whether our system will solve the contact-tracing problem, but we believe it is a technically sound alternative worth exploring. In addition, it serves the main purpose of this paper: to encourage the research community to revisit the design of digital contact-tracing solutions that are efficient to mitigate future waves of COVID-19 pandemic and if not, at least, future epidemics. 

\section{Solution Rationale}

We propose a novel contact-tracing solution that uses geolocation data of billions of users to find people that have been in contact to individuals tested positive. We refer to them as \emph{risk contacts}. The geolocation information is owned by BigTech companies referred to as Location Providers (LPs) in this paper, and the information of users tested positive is owned by Health Authorities (HA). The core of our solution can be described as follow: HAs send to LPs the IDs of infected users. LPs use the location information they own to find the risk contacts of the received IDs (according to the guidelines provided by epidemiology experts) and send back the list of risk contacts IDs to the HA. Finally, HAs reach out the risk contacts to inform them about the prevention protocol they have to follow. 

Note that for practical purposes, we propose to use the mobile phone number of individuals as user IDs in our solution. LPs know the mobile phone number of a major part of the users using their services, and it is reasonable to assume HAs record the mobile phone of infected users to communicate with them. 


Unfortunately, the direct exchange of data in clear between HAs and LPs presents important privacy issues. 
In particular, LPs should not receive in clear IDs of infected individuals and HAs should not be able to link the IDs of risk contacts to their correspondent infected user. 
Our solution addresses this challenge allowing to perform the contact-tracing task with strong privacy guarantees. To this end, we define an architecture and a communication protocol that involve in addition to LPs and HAs two more players: an Identity Provider (IDP) and an Independent Third-Party Authority (ITPA).

\subsection{Why using geolocation data?}

\noindent \textbf{Adoption:} The main limitation of contact-tracing based on mobile apps is the need to achieve a high rate of active users. This is a major bottleneck that so far has led every attempt in this line to fail.

Our solution avoids this bottleneck by using large-scale already available geolocation data owned by BigTech companies. To explicitly compare the penetration of BigTechs' data vs. BT mobile apps,  Table \ref{tab:penetrations_mobiles} shows for 18 countries we have found data on the number of installations of contact tracing apps: 1) the penetration rate of smartphones, Android OS \cite{Dimoco}\cite{statcounter}\cite{pew} and the Monthly Active Users (MAU) reported by FB;\footnote{\url{https://developers.facebook.com/docs/marketing-apis}} 2) the penetration rate of BT mobile-app in number of installations as well as an estimation of its penetration in terms of active users. The list of sources we have used to report the number of mobile apps installations can be accessed here.\footnote{\url{https://fdvt.org/files/COVID_APPS_SOURCES.pdf}} Note that, to the best of our knowledge, Switzerland is the unique country reporting the percentage of active users of its app, 63\% as of Dec 21, 2020 \cite{swiss_stats_app} . In order to have an estimation of the fraction of active users for other countries reporting the number of installations, we apply to their number of installations the Swiss ratio. 


    
According to our estimation, none of the countries reach a significant adoption rate close to 40\% for the contact-tracing mobile apps, and only 5 countries are above 20\%. In contrast, Facebook penetration is beyond 50\% in all countries but Germany (45.5\%). Similarly, the penetration of Android is higher than 40\% in all countries but US (32\%) and Switzerland (39\%). Note the Android penetration just represents a lower bound of Google penetration. Google has few other extremely popular apps such as Gmail and Google Maps that are widely used by iOS users.
\vspace{0.2cm}

\noindent \textbf{Accuracy:} BigTech companies use sophisticated techniques combining GPS, WiFi and cellular networks signals to geolocate users with high precision both outdoors and indoors \cite{GPSAccuracy}\cite{WifiIndoorGoogle}.  Indeed, Google claims to be able to geolocate users with an accuracy of 1 to 2 meters using multilateration algorithms based on the WiFi signal from 3 access points. \cite{WifiIndoorGoogle}. 
\vspace{0.2cm}

Therefore, high penetration rates and location accuracy of BigTechs present them as a data source that may be enough to implement efficient contact-tracing solutions. Recent research works, which uses data from LPs with much lower penetration than FB or Google,  also backup this hypothesis \cite{Aleta2020.12.15.20248273}.

\begin{table*}[t]
\centering
\small
\begin{tabular}{|l|c|c|c|c|c|}
\hline
\multicolumn{1}{|c|}{\multirow{2}{*}{\textbf{Country}}} & \multicolumn{1}{c|}{\multirow{2}{*}{\textbf{Smartphone}}} & \multicolumn{1}{c|}{\multirow{2}{*}{\textbf{Android}}} & \multicolumn{1}{c|}{\multirow{2}{*}{\textbf{Facebook}}} & \multicolumn{2}{c|}{\textbf{BT mobile apps}} \\ \cline{5-6} 
\multicolumn{1}{|c|}{} & \multicolumn{1}{c|}{} & \multicolumn{1}{c|}{} & \multicolumn{1}{c|}{} & \multicolumn{1}{c|}{\textbf{Installations}} & \multicolumn{1}{c|}{\textbf{\begin{tabular}[c]{@{}c@{}}Estimated\\ active users\end{tabular}}} \\ \hline
Australia & 105 & 44 & 71.42 & 27.6 & 17.4 \\ \hline
Austria & 117 & 78 & 50.25 & 9 & 5.7 \\ \hline
Belgium & 68 & 41 & 65.00 & 12.2 & 7.7 \\ \hline
Croatia & 71 & 59 & 50.84 & 2 & 1,3 \\ \hline
Czech Rep & 84 & 66 & 53.32 & 14 & 8.8 \\ \hline
Denmark & 115 & 55 & 71.03 & 34.8 & 21.9 \\ \hline
Finland & 140 & 97 & 59.65 & 45.3 & 28.5 \\ \hline
France & 79 & 51 & 58.35 & 9.5 & 6 \\ \hline
Germany & 90 & 61 & 45.50 & 34.5 & 21.7 \\ \hline
Ireland & 78 & 42 & 65.54 & 40.5 & 25.5 \\ \hline
Italy & 84 & 62 & 57.80 & 21.1 & 13.3 \\ \hline
Latvia & 96 & 69 & 52.45 & 9.1 & 5.7 \\ \hline
Netherlands & 82 & 48 & 63.09 & 25 & 15.8 \\ \hline
Portugal & 104 & 78 & 67.47 & 1 & 0.6 \\ \hline
Spain & 90 & 71 & 62.05 & 11.5 & 7.2 \\ \hline
Switzerland & 97 & 39 & 52.38 & 33.4 & 21.1 \\ \hline
United Kingdom & 85 & 40 & 66.64 & 23.8 & 15,1 \\ \hline
United States & 81 & 32 & 69.90 & 2.5 & 1.6 \\ \hline
\end{tabular}%
\caption{Penetration in percentage of smartphones, Android, Facebook and contact-tracing apps installations and estimated active users for 18 countries. The population of each country to compute the penetration is obtained from the World Bank Database \cite{wb_data}.}
\label{tab:penetrations_mobiles}
\end{table*}


\subsection{Other benefits}

The proposed solution allows to monitor its performance. In addition, geographical locations can be associated to specific categories referred to as Point of Interests (POIs). For instance, a given location can be mapped to a restaurant, a train station or a hospital. 
Our solution exploits this to provide statistical distribution of the POIs visited by infected users vs. POIs visited by the general population. The comparison of these distributions may help to identify statistical biases in POIs regularly more visited by infected users, that might be infection hotspots.

\subsection{Privacy requirements}
\label{subsec:privacy_requirements}

On the one hand, privacy experts and Data Protection Authorities (DPAs) have shown concerns to use geolocation information for digital contact tracing. They basically argue that it may ease governments through their HAs to implement massive surveillance due to the scalability provided by digital technologies. Therefore, \emph{our solution should limit the ability of HAs to massively infer the contact graph information of individuals using the data received from LPs. In addition, it should provide privacy provisions to allow revealing targeted attacks willing to infer the contact graph of particular individuals.}

On the other hand, BigTech companies have means to infer the identity of infected individuals. They can leverage geolocation data but also other information sources such as emails, posts in social networks or queries in search engines they own. For instance, they can detect a user who visited a testing facility after visiting its website and then remains at home for a period similar to the mandatory quarantine period. Therefore, we believe that proposals, as ours, leveraging BigTech companies' geolocation data do not impose any extra risk to infected users' privacy. In spite of this, appropriate privacy guarantees should be provided. In particular, \emph{our solution should not provide LPs explicit information about the identity of infected users. It also should limit the ability of LPs to infer such identities from the information received from HAs.}


\subsection{Meeting Privacy Requirements}
\label{subsec:meeting_privacy}

In order to meet the defined privacy requirements we leverage the following principles: K-anonymity, basic cryptography and non-repudiation auditing.

\noindent \textbf{K-anonymity}: In our solution the HA sends a list of user IDs to the LP and the LP answers with the risk contacts of those user IDs. 

Leveraging K-anonymity principles, the HA mixes in its request  $M$ IDs from infected users and $N$ real random IDs (i.e., random mobile phone numbers associated to real users) where $M <<< N$. This serves to anonymize the identity of infected users and to hinder the capacity of LPs to easily infer the IDs belonging to infected users. The random IDs used by the HA are provided by the Identity Provider (IDP) in order to guarantee that they are existing IDs. In our solution, IDPs are represented by mobile network operators. 

In addition, the HA must aggregate the IDs into groups. There are two types of groups: \emph{infected groups} include exclusively IDs from infected users; \emph{random groups} include IDs from random users or a mix of random and infected users.  The messages from the HA to the LP include $K$ groups from which only $L$ are infected groups, where $L <<< K$. Upon the reception of a request message from the HA, the LP computes the risk contacts of each user ID. After that, it aggregates together in the reply the risk contacts of all user IDs in a single group. This aggregation process relies on the K-anonymity concept to prevent the HA from linking the received risk contacts IDs to a specific individual. Note that the larger is the size of the groups the higher are the privacy guarantees.


\noindent \textbf{Cryptography:} A honest HA is interested only on the risk contacts IDs associated to infected groups. To hinder the ability of HAs to access contacts IDs from random groups, the LP encrypts the list of contacts of each group (included in the reply to the HA) using a different key per group. Therefore, the HA receives the contact IDs of all groups encrypted. To retrieve the keys of infected groups, the HA has to send a request to an intermediary that we refer to as Independent Third-Party Authority (ITPA). In this request, the HA indicates the total number of groups in the query as well as the ID of infected groups. In turn, the ITPA requests the keys of all groups to the LP and forwards to the HA only the keys associated to the infected groups. Finally, using the received keys, the HA obtains the risk contact IDs associated only to infected groups, thus completing the contact tracing procedure.

\noindent \textbf{Non-repudiation auditing:} Our solution relies on the concept of liability to guarantee privacy rights of the users. Note that this is a widely adopted approach in the legal system of advanced democracies. For instance, a state cannot prevent anyone from driving above the speed limit, but anyone doing so is liable for it. In the case of privacy, a state cannot prevent a BigTech company implementing privacy intrusive practices, but punish them in case an auditing process reveals the use of those practices. Therefore, a HA or a LP that uses the data they receive for purposes different than contact-tracing will be liable for it.

For instance, a malicious HA can implement a targeted attack (see Section \ref{sec:attacks}) to unveil the contact graph of an individual and leaking it to other government branches. This would be a crime equivalent to leaking the medical record of a target individual to other government branches. Our solution collects the required non-repudiation proofs to be used by the corresponding auditing entity to unveil any potential attack by a HA.



\begin{figure*}[t]
  \centering
  \includegraphics[width=\linewidth]{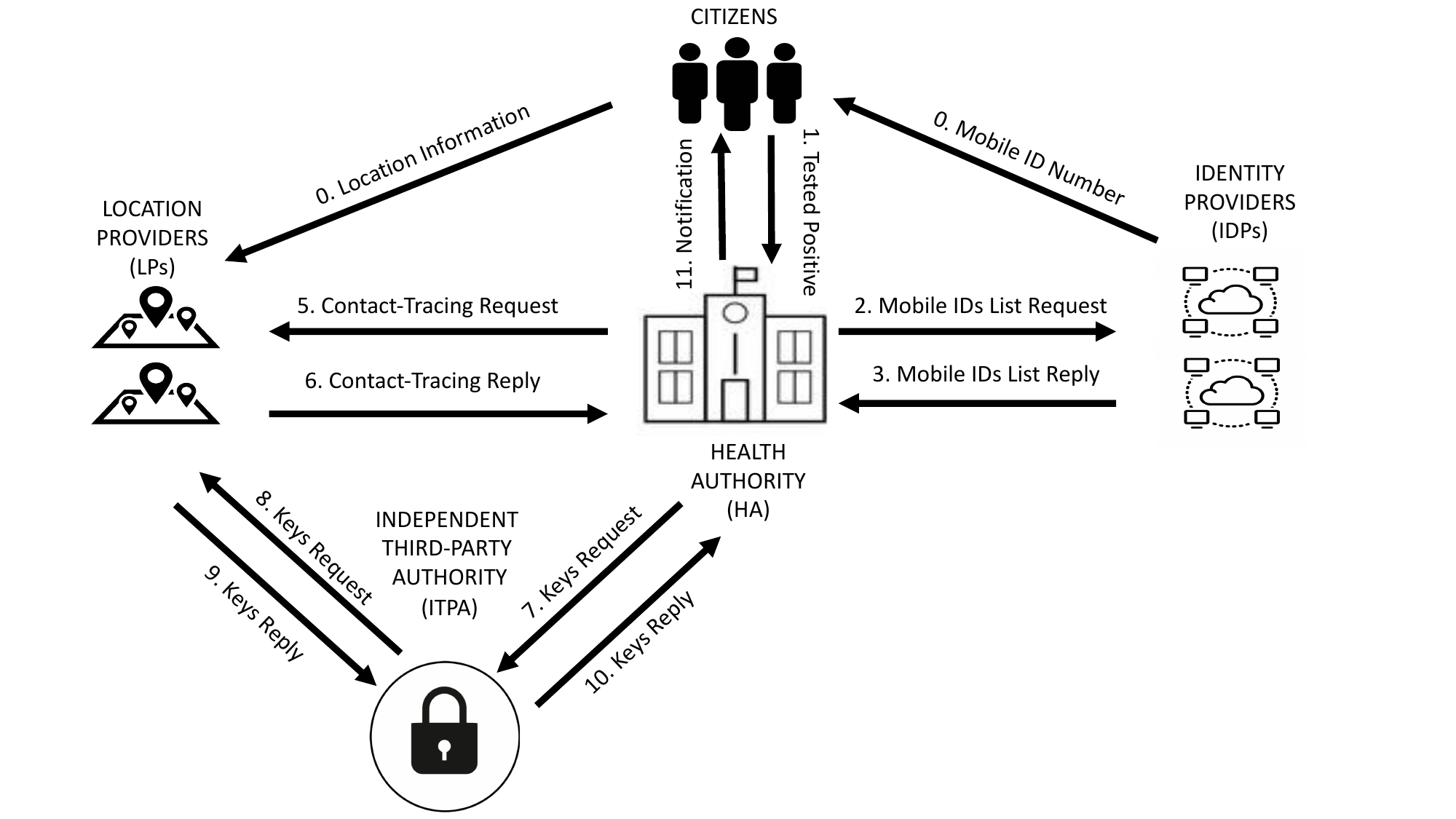}
  \caption{Proposed contact-tracing protocol and architecture.}
  \label{fig:protocol}
\end{figure*}

\section{Protocol for contact-tracing using Location Providers information}

In this section we describe the steps of the communication protocol including the sequence of messages exchanged by the four players involved in our solution: Health Authority (HA), Location Provider (LP), Identity Provider (IDP) and Independent Third Party Authority (ITPA). 

\noindent \textbf{Step 0}: This step refers to the basic context our solution relies on. On the one hand, LPs record historical location information from users running their OSes, mobile apps, etc. In addition, they also store the mobile phone number for a major portion of the users. On the other hand, IDPs (i.e., mobile operators) provide users with mobile phone numbers that serve as user IDs in our solution.

\noindent \textbf{Step 1}: The HA obtains the IDs of users that have been tested positive in a given time window (e.g., a day).

\noindent \textbf{Step 2}: The HA triggers the contact-tracing process by requesting the IDP a list of $N$ user IDs (i.e., real mobile phone numbers). The value of $N$ is decided by the HA and may differ from one request to another. 

There are few remarks to be considered: (1) This message includes a unique identifier referred to as \emph{Transaction ID} that will be included in all the remaining messages in the process; (2) The message is signed with the private key of the HA. Note that in the rest of the process all entities sign with their private key the messages they send.

\noindent \textbf{Step 3}: The IDP responds the HA request with a a list of $N$ random user IDs. 

\noindent \textbf{Step 4}: The HA creates $K$ groups. As explained above, only $L$ of these groups are infected groups and $K-L$ are random groups. The resulting groups are included in a \emph{Contact-Tracing Request} message that is sent to the LP. It is important to note that  the user IDs included in an infected group cannot be present in other infected groups neither in this request nor in past or future requests. 

\noindent \textbf{Step 5}: Upon the reception of the \emph{Contact-Tracing Request} the LP runs the contact-tracing algorithm to identify the risk contact IDs of each user ID included in the request. The risk contact IDs from all users in a group are aggregated so that any link between a user ID and a risk contact ID is eliminated. 

In addition the LP collects the POIs visited by each user ID in a defined time window in the past (e.g., last 10 days). Then, it computes the distribution of types of POIs visited by the user IDs included in each group as well as the overall distribution of types of POIs visited by all user IDs included in the request.

The information associated to each group, i.e., list of risk contact IDs and distribution of type of POIs is encrypted with an independent key per group. 

Finally, the LP aggregates the encrypted information per group along with the distribution of types of POIs for all users IDs and creates a \emph{Contact-Tracing Reply} message that is sent to the HA. 

Three important remarks to consider are: (1) The LP must keep record of the key used to encrypt each group; (2) The contact tracing algorithm implemented by the LP as well as the number of days for the identification of visited POIs must be defined by epidemiologists and it is out of the scope of this paper; (3) the LP stores all the \emph{Contact-Tracing Request} messages received for auditing purposes.

\noindent \textbf{Step 6:} Upon the reception of the \emph{Contact-Tracing Reply} the HA needs to decrypt the information associated to infected groups, i.e., the risk contacts list and type of POIs distribution. To this end, it sends a \emph{Keys Request} message to the ITPA including the total number of groups included in the \emph{Contact-Tracing Request} and the identifiers of the infected groups. 

\noindent \textbf{Step 7:} The ITPA sends to the LP the \emph{Keys Request} message but it only includes the \emph{Transaction ID}. 

\noindent \textbf{Step 8:} Upon the reception of the \emph{Keys Request} message, the LP sends to the ITPA a \emph{Keys Reply} message including the keys for all groups.

\noindent \textbf{Step 9:} The ITPA checks if the number of keys in the received reply matches the actual number of groups reported by the HA. If the numbers are the same, the ITPA generates a \emph{Keys Reply} message to the HA that only includes the keys of the infected groups. Otherwise, the \emph{Keys Reply} message includes an error indicating that the reported number of groups does not match with the number of keys provide by the LP.

\noindent \textbf{Step 10:} Upon the reception of the \emph{Keys Reply} message, the HA decrypts the information about risk contacts and type of POIs distributions included in the \emph{Contact-Tracing Reply} for the groups of infected users.

\noindent \textbf{Step 11:} The HA gets in touch with risk contacts.

\section{Potential attacks and countermeasures}
\label{sec:attacks}
As explained above, our solution is designed to hinder both, LPs and HAs, from misbehaving to get access to information they are not authorized to obtain. Next, we explain in detail the countermeasures provided by our solution to avoid: (i) LPs trying to infer the IDs associated with infected individuals, (ii) HAs trying to obtain the contact graph of citizens.

\subsection{LP inference of infected users identity}

A malicious LP may intend to unveil the identity of infected users based on the information received in \emph{Contact-Tracing Request} messages (known as re-identification attack). To this end they could use a single request or combine subsequent requests to obtain the identity of infected users. 

In order to prevent re-identification attacks, the HA has to reuse the IDs that have been already used including them in random groups of subsequent requests. Otherwise, if random IDs are only used once and discarded, the LP could infer with very high probability that repeated IDs in different queries belong to infected individuals. 


In addition to reusing IDs, our solution relies on the K-anonymity principle. The number of random IDs, $N$, in request messages is several times larger than the number of infected user IDs, $M$. The complexity to perform a re-identification attack grows with the ratio $\frac{N}{M}$. In addition, our solution allows introducing a high level of randomness in the request messages to avoid that LPs can infer patterns that allow identifying groups including infected users IDs: (i) the number of infected and random user IDs differs from message to message, (ii) the number of groups in a message differs from message to message, (iii) the length of the different groups within the same message should also differ. In addition, the HA could send messages that do not include any infected user ID from time to time. 

Beyond the technical measures, the main argument to support our solution is that powerful LPs such as Google or Facebook willing to identify infected citizens can easily do it already with the information they own. Therefore, the privacy measures adopted in our solution provide sufficient guarantees to avoid increasing the risk of a potential re-identification attack by LPs.

\subsection{HAs inference of the contact graph of a user-id}

Our solution cannot prevent beforehand a malicious HA from obtaining the  contact graph of a particular individual. For instance, a HA can perform a targeted attack by using twice the same ID in two different infected groups (despite it is forbidden in our solution). The common risk contacts in the two groups may reveal the contact graph of the targeted individual. 


However, our solution keeps the required non-repudiation proofs to show such an attack has happened. The auditing entity just needs to check whether the HA has used the same ID twice (or more times) in groups of infected users in the same or different messages. The auditing entity can retrieve all the \emph{Contact-Tracing Request} messages from the LP. Similarly, the auditing entity retrieves from the ITPA, for each \emph{Contact-Tracing Request} message, what are the infected groups declared by the HA. With that information the auditing entity can easily identify attacks from the HA. The described auditing capacity provides privacy guarantees based on undeniable liability, a widely used technique in developed democracies. 

Finally, our recommendation is to run the described auditing process once a day to detect any malicious HA soon after it has implemented an attack. 

\section{Conclusion}

The only digital contact-tracing approach used so far to fight COVID-19 pandemic consists on the utilization of mobile apps that leverage Bluetooth technology to identify proximity encounters. The lack of sufficient adoption of such mobile-apps has led every single attempt in this direction to fail. 

Due to the importance that digital contact-tracing solutions may have to help fighting the pandemic, it is the obligation of researchers, public health authorities and technology companies to explore alternatives until an effective contact-tracing solution is found. To trigger this exploration effort, in this paper we propose a first alternative solution. We propose to use already existing scalable and accurate geolocation data, which is likely to serve to build an efficient digital contact-tracing solution. Our proposal defines an architecture that leverages such data and provide sufficient privacy guarantees to citizens.




\section*{Acknowledgments}
The research leading to these results received funding from the European Union’s Horizon 2020 innovation action programme under the grant agreement No 871370 (PIMCITY project); the Ministerio de Economía, Industria y Competitividad, Spain, and the European Social Fund(EU), under the Ramón y Cajal programme (Grant RyC-2015-17732); the Ministerio de Educación, Cultura y Deporte, Spain, through the FPU programme (Grant FPU16/05852); the Ministerio de Ciencia e Innovación under the project ACHILLES (Grant PID2019-104207RB-I00); the Community of Madrid synergic project EMPATIA-CM (Grant Y2018/TCS-5046); the Fundación BBVA under the project AERIS; and the NSERC Discovery Grant 2016-04521; and the Community of Madrid and Universidad Carlos III de Madrid for the funding of research projects on SARS-CoV-2 and COVID-19 disease, project name ``Multi-source and multi-method prediction to support COVID-19 policy decision making'', which was supported with REACT-EU funds from the European regional development fund ``a way of making Europe''.

\bibliographystyle{IEEEtran}
\bibliography{bibliography}





\end{document}